\documentstyle[11pt]{article}

\newtheorem{claim}{Claim}

\newtheorem{observation}{Observation}

\newcommand{\eq}[1]{(\ref{#1})}

\newcommand{\be}{\begin{equation}}
\newcommand{\ee}{\end{equation}}
\newcommand{\eps}{\varepsilon}

\newcommand{\calN}{{\cal N}}

\begin{document}
\thispagestyle{empty}
\setcounter{page}{1}
\setlength{\baselineskip}{1.5\baselineskip}

\title{Coding for the Feedback Gel'fand--Pinsker 
Channel and the Feedforward Wyner--Ziv Source}
\author{Neri Merhav\thanks{Department of Electrical Engineering,
Technion -- Israel Institute of Technology, Haifa 32000, Israel.
E--mail: {\tt merhav@ee.technion.ac.il}} 
\and
Tsachy Weissman\thanks{
Department of Electrical Engineering,
Stanford University,
Stanford, CA 94305-9510, USA.
Email: {\tt tsachy@stanford.edu}}}
\maketitle

\begin{abstract}
We consider both channel coding and source coding, with perfect
past feedback/feedforward, in the presence of side information. It
is first observed that feedback does not increase the capacity of
the Gel'fand--Pinsker channel, nor does feedforward  improve the
achievable rate-distortion performance in the Wyner-Ziv problem.
We then focus on the Gaussian case showing that, as in the absence
of side information, feedback/feedforward allows to efficiently
attain the respective performance limits. In particular, we derive
schemes via variations on that of Schalkwijk and Kailath. These
variants, which are as simple as their origin and require no
binning, are
shown to achieve, respectively,  the capacity of Costa's channel, and the
Wyner-Ziv  rate distortion function. Finally, we consider
the finite-alphabet setting and
derive schemes for both the channel and the source coding problems
that attain the fundamental limits, using variations on schemes of
Ahlswede and Ooi and Wornell, and of Martinian and Wornell,
respectively.
\end{abstract}
{\bf Index Terms:} Side information, feedback, feedforward, dirty paper, source--channel
duality.

\clearpage
\section{Introduction}
That feedback does not increase the capacity of a memoryless
channel, yet can dramatically simplify the schemes for achieving
it, is a well known fact (cf.\ \cite{OoiWornell1998} and the
literature survey therein). More recently, an analogous phenomenon
was shown to hold for the dual problem of lossy source coding with
perfect past feedback, aka `feedforward', at the decoder
\cite{WeissmanMerhavCOmpetpred03, {PradhanISIT04},
{MartinianWornell2004}}, a problem arising in contexts as diverse
as prediction theory, remote sensing, and control.

In this work, we revisit these problems to accommodate the
presence of side information. As is the case for problems without
feedback/feedforward, the only scenarios with fundamental limits,
and achieving schemes, that are not directly implied  from those
known for the absence of side information are, respectively, the
presence of side information only at the encoder, and only at the
decoder, for the channel coding and the source coding problems.

Our first observation in this context is that the fact that
feedback/feedforward does not improve the fundamental performance
limits carries over to these cases where side information is
present. To see this, consider first channel coding for the
Gel'fand--Pinsker channel \cite{GelfandPinsker1980} with feedback,
by which we mean the following: The channel state information
$S^n$ is available to the sender, and the memoryless channel has
transition probability $p(y|x,s)$ that depends on the input $X$
and the state $S$. $S_i$ are assumed i.i.d.$\sim p(s)$. For a
message index $W \in \{1,2,\ldots,2^{nR}\}$, the $i$-th channel input is of the
form $X_i (W, S^n, Y^{i-1})$, i.e., allowed to depend on the past
channel output symbols. Decoding is, as usual, based on the
channel output $Y^n$.
\begin{observation} \label{observation for GP channel}
\emph{Feedback does not increase the capacity of the
Gel'fand--Pinsker channel.}
\end{observation}
\emph{Proof:} One need merely observe that the original converse
proof of Gel'fand and Pinsker \cite{GelfandPinsker1980} is general
enough so as to include feedback. In other words, $U_i - (X_i,
S_i) - Y_i$, where $U_i = (W, S_{i+1}^n, Y^{i-1})$, continues to
be a Markov chain even in the presence of feedback. To see this,
note that $P(y_i|w, s^n, y^{i-1}) =  p(y_i| x_i, s_i)$ so $(W,
S^n, Y^{i-1}) - (X_i, S_i) - Y_i$ is a Markov chain and therefore,
since $U_i$ is a deterministic function of $(W, S^n, Y^{i-1})$, so
is $U_i - (X_i, S_i) - Y_i$. Q.E.D. 

Though our interest
in this work, and the schemes we develop, are for the case of
non-causal state information, we mention in passing that a similar
conclusion applies also for the Shannon channel with causal SI,
where the $i$-th channel input is of the form $X_i (W, S^i,
Y^{i-1})$. The independence between $U_i$ and $S_i$ in the causal
case is readily verified to persevere in the presence of feedback,
implying: \emph{Feedback does not increase the capacity of the
Shannon channel (with causal SI).}

Moving to the source coding analogue, consider the  problem of
Wyner--Ziv source coding \cite{WynerZiv1976} with feedforward: The
source and side information are generated as independent drawings
of the pair $( X_i, Y_i )$. Encoding, as in the original problem,
is done by mapping the sequence $X^n$ into  $T \in \{1,2,\ldots,2^{nR}\}$. The
$i$-th reconstruction this time is of the form $\hat{X}_i (T, Y^n,
X^{i-1})$, i.e., allowed to depend also on the past,
non-quantized, past source symbols. This setting is the extension
of the source coding with feedforward problem
\cite{WeissmanMerhavCOmpetpred03, {PradhanISIT04},
{MartinianWornell2004}} to the case of side information at the
decoder.
\begin{observation} \label{observation for WZ problem}
\emph{Feedforward does not improve the rate distortion tradeoff in
the Wyner--Ziv problem.}
\end{observation}
\emph{Proof:} Here too, the original converse proof carries over
essentially unchanged. Specifically, in the notation of
\cite[Section 14.9]{CoverThomas1991}, we only need to add
$X^{i-1}$ to $W_i$, resulting in $W_i = (T, Y^{i-1}, Y_{i+1}^n,
X^{i-1} )$. The converse proof of \cite[Theorem
14.9.1]{CoverThomas1991} carries over verbatim (erasing line
(14.298) therein), since $W_i - X_i - Y_i$ continues to form a
Markov chain under this modified $W_i$. Q.E.D.

Given observations \ref{observation for GP channel} and
\ref{observation for WZ problem}, it is natural to ask whether,
similarly as in the absence of side information,
feedback/feedforward can lead to simple  schemes for attaining the
fundamental limits. For the Gaussian case, we answer this question
in the affirmative in the next section. More specifically,  we
present efficient schemes that exploit feedback/feedforward to
achieve the capacity of Costa's channel \cite{Costa83}, and the
Wyner--Ziv function  for a source which is a
Gaussian-noise-corrupted version of the side information. Our
schemes, which  are variations on those of Schalkwijk and Kailath
\cite{SchalkwijkKailath1966, {Schalkwijk1966}}, are as efficient
as their origin and, in particular, do not require binning. In
Section \ref{Finite alphabets}, we consider the finite alphabet
setting and derive a scheme for the Gel'fand--Pinsker channel with
feedback, building on the ideas of \cite{Ahlswede71,
{OoiWornell1998}}. We also derive a scheme for the dual
problem of Wyner--Ziv coding with feedforward, by extending the
approach of \cite{MartinianWornell2004}. Our schemes for the
finite--alphabet setting rely on Slepian--Wolf coding \cite{SlepianWolf73},
and thus we make no claim at this point regarding the efficiency
with which they can be implemented (in comparison to the efficiency of
practical schemes for the Gel'fand--Pinsker channel and the
Wyner--Ziv problem in the absence of feedback/feedforward). They are, however,
conceptually simple and suggest another view on the information--theoretic
formulas of the Gel'fand--Pinsker capacity and the Wyner--Ziv rate--distortion
function. They also shed light on yet another aspect of the duality between
source coding and channel coding with side information.

\section{Variations on the  Schalkwijk--Kailath Schemes}

\subsection{Writing on Dirty Paper 
with On--line Proofreading: Costa's Channel with Feedback}

Consider the channel $Y_i=X_i+S_i+Z_i,$ where $\{S_i\}$ is an
interference signal (with $ES_i=0$ and $\sigma_S^2=ES_i^2 <
\infty$) known to the encoder, and $\{Z_i\}$ is zero--mean,
i.i.d.\ Gaussian noise with variance $\sigma_Z^2$. Let the
transmission power be limited to $P$. We now describe a modified
version of the scheme of \cite{Schalkwijk1966} for coding with
feedback, which achieves the capacity $C=\frac{1}{2}\log
(1+P/\sigma_Z^2)$. Moreover, for every $R < C$, the error
probability is identical to that of the original scheme, as if
$S_i$ were identically zero, namely, it decays
double--exponentially rapidly with $C-R$.

\noindent \underline{Initialization:} Define
$\alpha=\sqrt{1+P/\sigma_Z^2}$, and $g=\sqrt{P/\sigma_Z^2}$. Given
a message $m=0,1,\ldots,M-1$, $M=2^{nR}$, let $\theta=(m+1/2)/M$.
Given $S^n=(S_1,\ldots,S_n)$ define $\psi_2=S_1/\alpha$, and for
$i=2,3,\ldots,n$, compute recursively:
$$\psi_{i+1}=\psi_i+\left(1-\frac{1}{\alpha^2}\right)\frac{S_i}{\alpha^{i-1}g}.$$
Finally, let $\theta'=\theta+\psi_{n+1}$.

\noindent \underline{Recursion:} For $i=1$, set $X_{1,1}=0.5$ and
transmit $\alpha(X_{1,1}-\theta')$. At the receiver, compute
$X_{2,1}=X_{1,2}=X_{1,1}-\frac{Y_1}{\alpha}$ and send $X_{2,1}$
back to the transmitter. For $i=2,3,\ldots,n$, transmit
$\alpha^{i-1}g(X_{i,1}-\theta'+\psi_i)$. At the receiver, compute
$X_{i,2}=X_{i,1}-\frac{Y_i}{\alpha^{i-1}g},$ then update
$$X_{(i+1),1}=\frac{1}{\alpha^2}X_{i,1}+\left(1-\frac{1}{\alpha^2}\right)X_{i,2},$$
and (for $i < n$) send $X_{(i+1),1}$ back to the transmitter.\\
Finally, decode $m$ by quantizing $X_{(n+1),1}$ to its message
interval.

\noindent \underline{Analysis}: First, note that
\begin{eqnarray}
X_{2,1} &=&
X_{1,1}-\frac{\alpha(X_{1,1}-\theta')+S_1+Z_1}{\alpha}\nonumber\\
&=& \theta'-\frac{S_1}{\alpha}-\frac{Z_1}{\alpha}\nonumber\\
&=&\theta'-\psi_2-\frac{Z_1}{\alpha}. \label{init}
\end{eqnarray}
We now argue that for all $i\ge 2$,
$X_{i,1}=\theta'-\psi_i-\phi_i$, where $\{\psi_i\}$ are defined as
above, and $\{\phi_i\}$ are defined by $\phi_2=Z_1/\alpha$ and by
the recursion
$$\phi_{i+1}=\frac{1}{\alpha^2}\phi_i+
\left(1-\frac{1}{\alpha^2}\right)\frac{Z_i}{\alpha^{i-1}g},~~~~i=2,3,\ldots,n.$$
We prove  this by induction: For $i=2$, this has been shown
already in eq.\ (\ref{init}). Assuming now that the hypothesis is
true for a given $i\ge 2$, then
\begin{eqnarray}
X_{(i+1),1}&=&\frac{1}{\alpha^2}X_{i,1}+\left(1-\frac{1}{\alpha^2}\right)X_{i,2}\nonumber\\
&=&\frac{1}{\alpha^2}(\theta'-\psi_i-\phi_i)+\left(1-\frac{1}{\alpha^2}\right)
\cdot \nonumber \\ 
& &\left[X_{i,1}-\frac{\alpha^{i-1}g(X_{i,1}-\theta'+\psi_i)+S_i+Z_i}{\alpha^{i-1}g}\right]\nonumber\\
&=&\frac{1}{\alpha^2}(\theta'-\psi_i-\phi_i)+\left(1-\frac{1}{\alpha^2}\right)
\left[\theta'-\psi_i-\frac{S_i+Z_i}{\alpha^{i-1}g}\right]\nonumber\\
&=&\theta'-\left[\psi_i+\left(1-\frac{1}{\alpha^2}\right)\frac{S_i}{\alpha^{i-1}g}\right]
\nonumber \\ 
& &-\left[\frac{1}{\alpha^2}\phi_i+\left(1-\frac{1}{\alpha^2}
\right)\frac{Z_i}{\alpha^{i-1}g}\right]\nonumber\\
&=&\theta'-\psi_{i+1}-\phi_{i+1},
\end{eqnarray}
confirming the induction hypothesis for $i+1$. Thus, for $i=n+1$,
we get
\begin{equation}
X_{(n+1),1}=\theta'-\psi_{n+1}-\phi_{n+1}=\theta-\phi_{n+1}. 
\end{equation}
But $\phi_{n+1}$ is exactly the estimation error variable in
\cite{Schalkwijk1966}, whose variance has been shown to be
$\sigma_Z^2/\alpha^{2n}$. Thus, the decision made by this scheme
is identical to that of Schalkwijk's scheme (with $S^n=0$) for
every realization of the noise sequence. Obviously, the error
performance is then the same too.

As for the transmission power, we will distinguish again between
$i=1$ and $i\ge 2$. For $i=1$, the transmission power is
approximately $\alpha^2(1/12+\mbox{Var}\{\psi_{n+1}\})$, where
$1/12$ approximates the variance of $\theta$ as one corresponding
to the uniform distribution in $[0,1]$, and
$\mbox{Var}\{\psi_{n+1}\}$ is bounded independently of $n$ since
$\psi_{n+1}$ is a linear combination of $\{S_i\}$ with
coefficients that decay exponentially with $i$. As for $i\ge 2$,
the transmission power is
\begin{eqnarray}
\alpha^{2(i-1)}g^2E(X_{i,1}-\theta'+\psi_i)^2&=&\alpha^{2(i-1)}g^2E\phi_i^2\nonumber\\
= \alpha^{2(i-1)}g^2\frac{\sigma_Z^2}{\alpha^{2(i-1)}}
&=&\sigma_Z^2g^2 =P,
\end{eqnarray}
where the second equality has been proved in \cite{Schalkwijk1966}
(and can also easily be seen by induction, using the recursive
definition of $\{\phi_i\}$). Thus, except for $i=1$, the
transmission power is $P$ at all times, which means that for large
$n$ the total average power tends to $P$.

At the point, a few comments are in order:
\begin{itemize}
\item[1.] We have seen that in the presence of feedback, it is
possible to achieve capacity with a simple scheme, without
binning. 
\item[2.] While in the absence of feedback \cite{Costa83},
the idea is not to `fight' the interference by trying to
pre--cancel it but rather to harness it to our own benefit, here
the pre--cancelling approach seems to be fruitful. This is
manifested both at the transmitter, where the contribution of
$\{S_i\}$ to the estimation error to be transmitted is cancelled
in order to save power, and in the definition of $\theta'$, which
shifts $\theta$ by an amount ($\psi_{n+1}$) which pre--cancels the
contribution of $\{S_i\}$ to the error of the final estimator.
\item[3.] As mentioned earlier, operatively, this scheme gives exactly
the same estimation and decoding as in \cite{Schalkwijk1966} for
every realization of the noise process, and as if $S^n$ were
non--existent ($S^n=0$). 
\item[4.] Similarly to the non--feedback
case, the probability law of $\{S_n\}$ is immaterial. The only
requirement is that $\sigma_S^2 < \infty$ to assure that the
expected power used at time $i=1$ is finite. 
\item[5.] Note that the
non--causal dependence of the transmission on $S^n$ is only via
one number, $\psi_{n+1}$.
\end{itemize}

\subsection{A Scheme for  Wyner--Ziv Coding with Feedforward}

Consider first rate distortion coding with feedforward in the
absence of side information \cite{WeissmanMerhavCOmpetpred03}. Let
$\{X_i\}_{i=1}^l$ be i.i.d.\ ${\cal N}(0, \sigma^2)$ and, 
for a given positive real $\beta$, let 
\begin{equation}
Y = -
\sum_{k=2}^l \sqrt{\beta^2 -1} \beta^{-(k+1)} X_k - \beta^{-1}
X_1. 
\end{equation}
Let $\hat{Y}$ be the quantized version of $Y$ using a
uniform scalar quantizer on the interval $[-\Delta/2, \Delta/2]$ with $M$
levels (truncating values outside the interval). Encoder describes
$\hat{Y}$ to decoder by giving index $I(\hat{Y})$ of the
quantization cell. Decoder reconstructs as follows: $\hat{X}_1 = -
\beta \hat{Y}$, $\hat{X}_2 = \sqrt{\beta^2 - 1} (\hat{X}_1 - X_1
)$, $\hat{X}_i = \beta \hat{X}_{i-1} - (\beta^2 - 1) \beta^{-1}
X_{i-1}$ for $i=3, \ldots, l$. It was shown in
\cite{PradhanISIT04} that,
 for $l \geq 1$,
\be \label{eq: identity for CP scheme} \frac{1}{l} \sum_{i=1}^l E
(X_i - \hat{X}_i)^2 = \frac{E(Y-\hat{Y})^2 \beta^{2 l}}{l} +
\frac{\sigma^2 (l \beta^2 - \beta^2)}{l \beta^4}.  \ee
To see how this scheme attains the rate distortion function, fix
the rate $R$ (so $M=2^{R l}$) and a small $\eps >0$ throughout.
Take $\beta = 2^{R- 2 \eps}$ and $\Delta = 2^{l \eps}$. We note
the following:
\begin{enumerate}
    \item $\sum_{k=2}^\infty (\beta^2 -1)
\beta^{- 2 (k+1)} < \infty$ so the variance of $Y$ is bounded
(does not exceed a fixed value) regardless of $l$.
    \item $\mbox{Pr}\{Y \not \in [-\Delta/2, \Delta/2]\}$ is diminishing with
    $l$ (in fact, double--exponentially rapidly 
    since $Y$ is Gaussian with bounded variance and is $\Delta$ exponentially growing with $l$).
    \item In $[-\Delta/2, \Delta/2]$ we are performing uniform
    quantization with resolution $\Delta/M = 2^{-(R - \eps ) l}$.
    \item The two previous items 
    imply that $E (Y-\hat{Y})^2 \leq c (\Delta /M )^2 = c 2^{- 2 (R - \eps )
    l}$ for an $l$-independent constant $c$ (in fact, a high--resolution 
    quantization argument will give the more refined
    $E (Y-\hat{Y})^2  \sim \frac{1}{12} 2^{- 2 (R - \eps )
    l}$).
    \item Substituting into \eq{eq: identity for CP scheme}, we
    get, as $l$ grows large, that the first term on the right side
    diminishes, while the second one converges to $\frac{\sigma^2 }{
    \beta^2} = \sigma^2 2^{-2 (R- 2 \eps)}$, which is the distortion--rate 
    function (up to the small $\eps$
    factor).
\end{enumerate}
Performance analysis for our scheme below will rely also on:
\begin{claim} \label{claim on robustness}
The scheme described is robust in the sense that if the decoder
 receives any index $\tilde{I}$ such that $\log |I( \hat{Y} ) - \tilde{I}| =
 o(l)$, then the distortion converges, as for the original scheme,
 to $\sigma^2 2^{-2 (R- 2 \eps)}$.
\end{claim}
\emph{Proof:} The distance between the centers of two
adjacent quantization cells is $2^{-(R-\eps) l}$, so, letting
$\tilde{Y}$ denote the value of $\hat{Y}$ that the decoder assumes
based on $\tilde{I}$, $|\hat{Y}-\tilde{Y}| \leq 2^{-(R-\eps +
o(1)) l}$. The error in reconstruction due to this discrepancy can
increase from one component to the next by a factor of $\beta =
2^{R- 2 \eps}$, so the overall distance between the reconstruction
based on $\tilde{I}$ and that based on $I$ is diminishing  (this
is why  $\beta = 2^{R-2 \eps}$ rather than $\beta = 2^{R- \eps}$
was taken). Q.E.D. 

Consider now the Wyner--Ziv
problem with perfect feedforword on the past source symbols at the
decoder. Assume:
\begin{enumerate}
    \item $\{Y_i\}$ is an arbitrarily distributed side-information signal available only at
    the decoder.
    \item $\{X_i\}$, the source signal, is given by $X_i = Y_i +
    N_i$, where $\{N_i\}$ is i.i.d.\ $\calN(0, \sigma^2)$,
    independent
    of $\{Y_i\}$.
\end{enumerate}
Consider next the following scheme for this setting:
\begin{itemize}
    \item Encoder: operate \emph{exactly} as encoder associated
    with \eq{eq: identity for CP scheme}.
    \item Decoder: \begin{enumerate} \item Add $
\sum_{k=2}^l \sqrt{\beta^2 -1} \beta^{-(k+1)} Y_k + \beta^{-1}
Y_1$ to received $\hat{Y}$. \item Input the result into the
decoder  using $\{N_i\}$ as the feedforward sequence (which is
possible since at time $i$  $X_{i-1}$ is revealed,  and  $Y_{i-1}$
is of course known).

 \item Let the reconstruction
be given by $\hat{X}_i = Y_i + \hat{N}_i$, where $\hat{N}_i$ is
output of the decoder from the previous stage.
\end{enumerate}
\end{itemize}
\begin{claim}
As $l \rightarrow \infty$, the distortion of the scheme described
converges to $\sigma^2 2^{-2 (R - 2 \eps)}$.
\end{claim}
\emph{Proof:} Since
 \begin{eqnarray*} \label{eq: Y in this case} Y & = &  - \sum_{k=2}^l
\sqrt{\beta^2 -1} \beta^{-(k+1)} X_k - \beta^{-1} X_1 \\ & = & -
\sum_{k=2}^l \sqrt{\beta^2 -1} \beta^{-(k+1)} Y_k \\ & & -
\beta^{-1} Y_1 - \sum_{k=2}^l \sqrt{\beta^2 -1} \beta^{-(k+1)} N_k
- \beta^{-1} N_1 , \end{eqnarray*}  assuming $Y \in [-\Delta/2,
\Delta /2]$, the quantization resolution implies $|Y - \hat{Y}|
\leq \Delta/M = 2^{-(R - \eps ) l}$, so the input (index) given to
the decoder  in the second stage is within $1$ from what it would
have received had encoding been performed (with scheme in \eq{eq:
identity for CP scheme}) directly on the $\{ N_i \}$ sequence.
Claim \ref{claim on robustness} implies then that the distortion
between $\{ N_i \}$ and $\{ \hat{N}_i \}$, hence also between $\{
X_i \}$ and $\{ \hat{X}_i \}$, is essentially $\sigma^2 2^{-2 (R-
2 \eps)}$. It only remains to argue that our assumption $Y \in
[-\Delta/2, \Delta /2]$ was justified. To this end, observe that:
\begin{eqnarray}\label{eq: bound on variance of Y}
    \mbox{Var}\{Y\} &\leq&   \mbox{Var} \left\{\sum_{k=2}^l 
    \sqrt{\beta^2 -1} \beta^{-(k+1)} Y_k + \beta^{-1} Y_1 \right\} \\ & &  + \sigma^2
    \left[ \sum_{k=2}^\infty ( \beta^2 -1 ) \beta^{-2 (k+1)}  + \beta^{-2}
    \right],
\end{eqnarray}
so as long as $\sum_{k=2}^l \sqrt{\beta^2 -1} \beta^{-(k+1)} Y_k +
\beta^{-1} Y_1$ has expectation and variance growing
sub-exponentially with $l$, which is the case for all but the
wildest processes, since $\Delta = 2^{l \eps}$, $\mbox{Pr}\{Y \in
[-\Delta/2, \Delta/2]\}$ is overwhelmingly small. Q.E.D.

Comments:
\begin{enumerate}
    \item The scheme is as simple as the channel coding one, with
    no binning required.
    \item This scheme achieves the conditional rate-distortion
    function for the case where the side information is available
    at both encoder and decoder, for an arbitrarily distributed
    side information process $\{Y_i\}$.
    \item Observation \ref{observation for WZ problem}, combined with the previous
    item,    implies that for the regular Wyner--Ziv problem, in the case
    where the pairs $(X_i,
    Y_i)$ are i.i.d., with $Y_i$ arbitrarily distributed and $X_i = Y_i +
    N_i$ for $N_i$ Gaussian and independent of $Y_i$, there is no
    loss due to the absence of side information at the encoder. This
     fact can be deduced also directly from the single-letter
     expression. Indeed, the argument used in \cite[Section 3]{Wyner1978} to
     show that the Wyner--Ziv function coincides with the
     conditional rate distortion function when $(X_i, Y_i)$ are
     jointly Gaussian is readily  seen to carry over to this
     more general case.

    \item The results of \cite{WeissmanMerhavCOmpetpred03} can be shown to imply, for an arbitrarily
    distributed SI process $\{Y_i\}$, and source given by $X_i = Y_i +
    N_i$, for i.i.d.\ (but arbitrarily distributed) process $\{ N_i
    \}$, that feedforward does not help for source coding with SI on
    both sides. Combined with the second item, this implies that
    the Wyner--Ziv performance in the presence of feedforward for an arbitrarily
    distributed SI process  and source given
    by $X_i = Y_i + N_i$, for $N_i$ i.i.d.\ Gaussian, coincides
    with that for SI at both sides. Furthermore, we have just shown a simple scheme attaining optimum performance for this case which is no less simple
    than had the SI been available at the encoder as well. Thus, not only is there no loss for not knowing the SI at the encoder in terms of the fundamental
    limit, there is also no loss in the simplicity of the scheme attaining
    it.
    \item Non-causal dependence of decoding on the SI in the above
    scheme is only once, in the first step, for computing $\sum_{k=2}^l \sqrt{\beta^2 -1} \beta^{-(k+1)} Y_k +
\beta^{-1} Y_1$. The reconstruction in the remaining steps uses
the SI causally.
\end{enumerate}

\section{Finite alphabets} \label{Finite alphabets}
\subsection{A Scheme for the Gel'fand--Pinsker Channel with Feedback}
Consider the finite-alphabet setting of the Gel'fand--Pinsker
channel, as described in the introduction. Let $S,U,X,Y$ have a
capacity--achieving distribution, namely, a distribution achieving
$\max_{p(u|s), f} [ I(U;Y) - I(U; S) ]$, where $X=f(U,S)$.
Consider the following scheme of coding with feedback for the
Gel'fand--Pinsker channel\footnote{Throughout, we ignore integer
constraints, writing, e.g., $ N/H(U|S)$ rather than $  \lceil
N/H(U|S) \rceil $.}, building on the ideas of \cite{Ahlswede71,
{OoiWornell1998}}:
\begin{itemize}
    \item \emph{Transmitter:}
        Maps the $N$ message bits into the sequence $U^{n_1}$, $n_1 =
        N/H(U|S)$, where $U^{n_1}$ is the output of the
        decoder corresponding to an optimal Slepian--Wolf encoder
        of $U^{n_1}$ for side information  $S^{n_1}$, when receiving the
        $N$ message bits as input from the encoder and observing
        the side information        $S^{n_1}$.

        \item   Sends $X^{n_1}$ through the channel, where $X_i = f (U_i,
        S_i)$, $1 \leq i \leq n_1$.

    \item \emph{Channel:}
        Corrupts $X^{n_1}$ according to $p(y|s,x)$.

    \item \emph{Receiver:}  Feeds channel output $Y^{n_1}$ back to the transmitter.

\item  \emph{Transmitter:}
        Using $Y^{n_1}$, compresses $U^{n_1}$ into $n_1 H(U|Y)$ new data
        bits.

        \item Maps these bits into the sequence $U_{n_1+1}^{n_1 + n_2}$, for $n_2 = n_1
        H(U|Y)/H(U|S)$,  by letting $U_{n_1+1}^{n_1 + n_2}$ be the output of the
        decoder corresponding to an optimal Slepian--Wolf encoder
        of $U$ for side information  $S$ (for $n_2$-tuples), when receiving the
        $n_1 H(U|Y)$ new data
        bits  as input from the encoder and observing
        the side information       $S_{n_1+1}^{n_1 + n_2}$.

        \item   Sends $X_{n_1+1}^{n_1 + n_2}$ through the channel, where $X_i = f (U_i,
        S_i)$, $n_1 + 1 \leq i \leq n_1 + n_2$.

    \item \emph{Channel:}
        Corrupts $X_{n_1+1}^{n_1 + n_2}$ according to $p(y|s,x)$.

    \item \emph{Receiver:}   Feeds channel output $Y_{n_1+1}^{n_1 + n_2}$ back to the transmitter.

    \item \emph{Transmitter:}  Using $Y_{n_1+1}^{n_1 + n_2}$, compresses $U_{n_1+1}^{n_1 + n_2}$ into $n_2 H(U|Y)$ new data
        bits, 
        \end{itemize}
and so on. After $k$ iterations of this process, letting $l_k =
\sum_{i=1}^{k} n_i$, use a simplistic termination code for
conveying the $n_k$-tuple
$
    U_{l_{k-1 } + 1 }^{ l_k}  $ to the decoder, allowed to be based
also on $ Y_{l_{k-1 } + 1}^{ l_k} $ that will be available from
the feedback. Thus, in effect, this termination code needs to
communicate $ \approx n_k H(U|Y) = N [ H(U|Y)/H(U|S) ]^k$
additional information bits.

\underline{Decoding:} Let $ \hat{U}_{l_{k-1} + 1 }^{ l_k } $
denote the decoder's estimated version of $U_{l_{k-1 } + 1 }^{
l_k}$, and let $\mathbf{b}_k$ be the binary $n_k H(U|S)$-tuple
obtained by taking the  output of the Slepian--Wolf encoder (used
at the $k$-th stage of the encoding) when this $n_k$-tuple is used
as its input. Let $\hat{U}_{l_{k-2} + 1 }^{ l_{k-1} } $ be the
conditional entropy decoding of an $n_{k-1}$-tuple of the source
$U$ given the corresponding $n_{k-1}$-tuple of $Y$ as side
information, for the $Y$ sequence $Y_{l_{k-2} + 1 }^{ l_{k-1} } $
and the binary encoding $\mathbf{b}_k$. Now feed the
$n_{k-1}$-tuple $\hat{U}_{l_{k-2} + 1 }^{ l_{k-1} } $ into the
Slepian--Wolf encoder used at the $k-1$-th stage of the encoding,
and let
 $\mathbf{b}_{k-1}$ be the binary
$n_{k-1} H(U|S)$-tuple obtained at its output. Continue this
process for $k$ iterations, until obtaining the binary $N$-tuple
$\mathbf{b}_{1}$, letting that be the decoded message bits.

\underline{Analysis:} The overall number of channel uses is
\begin{eqnarray*}\label{eq: num of channel uses}
    l_k + L  & = &  \frac{N}{H(U|S)} \sum_{i=1}^k \left[ H(U|Y)/H(U|S) \right]^{i-1} +
    L \\   & = & \frac{N}{H(U|S)} \frac{1 - \left[ H(U|Y)/H(U|S) \right]^{k}}{1- \left[ H(U|Y)/H(U|S) \right]}
    +L \\
      & \leq & \frac{N}{H(U|S) - H(U|Y)} + L ,
\end{eqnarray*}
where $L$ denotes the length of the termination code. In other
words, assuming $L \ll N$, the number of information bits per
channel use is essentially $H(U|S) - H(U|Y)= I(U;Y)-I(U;S)$, the
capacity. The probability of decoding error can readily be shown
to diminish, taking $k$  small enough so that the probability of
an error in the Slepian--Wolf coding at any one of the $k$ steps is
negligible, yet large enough so that the length $L$ of the
termination code required to reliably transmit the last block
(whose length decays exponentially with $k$) is negligible
relative to $N$.


\subsection{Wyner--Ziv Coding with Feedforward}
Assume the Wyner--Ziv setting where source and SI are i.i.d.\
drawings of $(X,Y)$. We further generate $U$ according to
$P_{U|X}$ (so  $U-X-Y$), and let $\hat{X} = f(U,Y)$, taking
$P_{U|X}$ and $f$ to be achievers of the Wyner--Ziv function.

\emph{Shaping Subsystem:}
 Given $x^n$ which
is $P_X$-typical, a shaper $S_{U|X} ( \cdot , x^n)$ is a 1-to-1
mapping from $\{0,1\}^{n H(U|X)}$ into $T_{U|X}[x^n]$ (where
$T_{U|X}[x^n]$ denotes the set of $u^n$-s that are jointly typical
with $x^n$). In other words, to every binary $n H(U|X)$-tuple $b$
there corresponds a (different) $u^n = S_{U|X} ( b , x^n)$ such
that $(u^n, x^n)$ are jointly typical. Let $S_{U|X}^{-1} ( \cdot ,
x^n)$ denote the inverse mapping of $S_{U|X} ( \cdot , x^n)$.
Existence of shapers follows from elementary facts known from the
method of types. Shaping systems can be implemented efficiently
via arithmetic coding \cite{OoiWornell1998,
{MartinianWornell2004}}.

\emph{Slepian--Wolf Coding:} Given a typical $u^n$, let $C_U (u^n)$
denote the bit sequence of length $n H(U|Y)$ resulting from an
essentially optimal Slepian--Wolf encoding of $u^n$ for the
presence of side information $Y^n$ at the decoder. For $b$ a
binary sequence of length $n H(U|Y)$ let  $C_U^{-1} (b, y^n)$
denote the reconstruction of the corresponding decoder when
receiving $b$ from the encoder and the side information sequence
is $y^n$.

\emph{Our Scheme:} Fixing $L$, $k$, we take the length of the
source sequence to be $n = L \sum_{j=0}^{k-1} \left[ H(U|Y)/H(U|X)
\right]^{j}$. The following scheme builds on the ideas in
\cite{MartinianWornell2004}.

\underline{Encoding:}
\begin{itemize}
    \item Initialize $T=1$, $l=L$, $j=1$, and reverse the input so that
     $X^n \rightarrow (X_n, X_{n-1}, \ldots, X_1)$
    \item Take the block of source samples $X^l$ and  generate a
    ``noisy version'' $U^l$ by passing $X^l$ through the
    ``channel'' $P_{U|X}$.
    \item \textbf{while} $j < k$ \textbf{do}:
    \item  Do Slepian--Wolf encoding of $U_T^{T+l}$ to obtain the
    binary $l H(U|Y)$-tuple $b = C_U (U_T^{T+l})$. Let
     $T = T + l + 1$,
     $l = L [H(U|Y)/H(U|X)]^j$,
   $U_{T}^{T+l} = S_{U|X} ( b , X_{T}^{T+l})$, and
    $j = j+1$
    \item \textbf{end while}
    \item \textbf{return} $b = C_U (U_T^{T+l}) = C_U (U_{n-l}^{n})$
\end{itemize}
\underline{Decoding:}
\begin{itemize}
    \item Initialize $T = n$, $j=k-1$.
 \item \textbf{while} $j \geq 0$ \textbf{do}:
    \item Let $l =  L [ H(U|Y)/H(U|X)
    ]^j $ and
      $ T = T - l $.
   Construct $ \hat{X}_{T}^{T+l} $ by letting $\hat{X}_i = f ( \hat{U}_i,
    Y_i)$ for each $T \leq i \leq T+l$, where $ \hat{U}_{T}^{T+l} = C_U^{-1} (b, Y_{T}^{T+l})
    $.
  Obtaining $X_T^{T+l}$ via the feedforward, let
 $b = S_{U|X}^{-1} ( U_{T}^{T+l} , X_T^{T+l}) $. Finally, let
 $j=j-1$

    \item \textbf{end while}

    \item \textbf{return}  the reversed version of $\hat{X}_1^n$

\end{itemize}

\emph{Performance:} For $k$ fixed and $L$ large the Slepian--Wolf
decoding is essentially error free, i.e., with high probability,
at each of the $k$ cycles of the \textbf{while} loop in the
decoding $\hat{U}_{T}^{T+l} = U_{T}^{T+l}$. Furthermore, the
reconstruction $ \hat{X}_{T}^{T+l} $ obtained at each of the $k$
cycles is, with high probability, jointly typical with
$X_{T}^{T+l} $. Thus the overall distortion is, with high
probability, approximately $E \rho (X, \hat{X})$. As for the rate
note that, by construction, the number of bits emitted by the
encoder is $L \left[ H(U|Y)/H(U|X) \right]^{k-1} \cdot H(U|Y)$,
while the number of source samples encoded is $n = L \frac{\left[
H(U|Y)/H(U|X)
    \right]^{k}-1}{\left[ H(U|Y)/H(U|X)
    \right] - 1}$. Thus essentially, for $1 \ll k \ll L$, the rate
    achieved is
$
    R \approx H(U|Y) - H(U|X) = I(U; X) - I(U;Y),
$
 the optimal Wyner--Ziv rate.
\section*{Acknowledgment}
N.\ Merhav would like to thank S.\ Shamai and Y.\ Steinberg for
interesting discussions in the early stages of this work.

\end{document}